\documentstyle[12pt]{article}

\begin{document}
\setcounter{page}{1}
\renewcommand{\thefootnote}{\fnsymbol{footnote}}
\pagestyle{plain} \vspace{1cm}
\begin{center}
\Large{\bf  Quantum Gravity and Recovery of Information in Black
Hole Evaporation}

\small \vspace{1cm} {\bf Kourosh Nozari$^{\rm a,b,}$\footnote{
knozari@umz.ac.ir}}\quad\quad and \quad\quad {\bf S. Hamid
Mehdipour$^{\rm
a,c,}$\footnote{h.mehdipour@umz.ac.ir}} \\
\vspace{0.5cm} {\it $^{a}$Department of Physics, Faculty of Basic
Sciences,
University of Mazandaran,\\
P. O. Box 47416-1467, Babolsar, IRAN}\\
{\it $^{b}$Research Institute for Astronomy and Astrophysics of
Maragha,
\\P. O. Box 55134-441, Maragha, IRAN}\\
{\it $^{c}$ Islamic Azad University, Lahijan Branch,\\
P. O. Box 1616, Lahijan, IRAN }\\

\end{center}
\vspace{1.5cm}
\begin{abstract}
The Generalized Uncertainty Principle (GUP), motivated by current
alternatives of quantum gravity, produces significant modifications
to the Hawking radiation and the final stage of black hole
evaporation. We show that incorporation of the GUP into the quantum
tunneling process (based on the null-geodesic method) causes
correlations between the tunneling probability of different modes in
the black hole radiation spectrum. In this manner, the quantum
information becomes encrypted in the Hawking radiation, and
information can be recovered as non-thermal GUP
correlations between tunneling probabilities of different modes.
\end{abstract}
\newpage
\section{Introduction}
Quantum gravity proposals commonly predict the existence of a
minimum observable length of the order of the Planck length [1-5].
Such a smallest length causes an alteration of the position-momentum
uncertainty relation [2] in such a way that $\Delta
x\geq\frac{1}{\Delta p}+\alpha L_{p}^{2}\Delta p$, where $\alpha$ is
a dimensionless constant of the order of unity that depends on the
details of the quantum gravity hypothesis. The above Generalized
Uncertainty Principle (GUP) can be obtained from string theory [3],
noncommutative quantum theory [4], loop quantum gravity [1], or from
black hole gedanken experiments [5]. In the standard limit, $\Delta
x\gg L_p$, it yields the ordinary uncertainty principle, $\Delta
x\Delta p\geq1$. The GUP becomes susceptible when the momentum and
distance scales are close to the Planck scale (see [6,7] and
references therein). For a spherical black hole the thermodynamic
quantities can be obtained in a heuristic manner using the standard
uncertainty principle [8]. Application of the GUP to black hole
thermodynamics modifies the results dramatically by incorporation of
quantum gravity effects in the final stages of evaporation with a
rich phenomenology [9] (see also [2,10,11]). Implications of the GUP
on various problems have been studied extensively some of which can
be obtained in Ref. [12].

Parikh and Wilczek [13] constructed a procedure to describe the
Hawking radiation emitted from a Schwarzschild black hole as a
tunneling process of a massless particle through its quantum
horizon. This procedure provides a leading correction to the
tunneling probability (emission rate) arising from reduction of the
black hole mass due to the energy carried by the emitted quantum.
However, due to its lack of correlations between the emission rates
of different modes in the black hole radiation spectrum the form of
the correction is not adequate by itself to return information. Here
we generalize the Parikh-Wilczek tunneling framework by the
incorporating quantum gravity effects that are manifested in the
existence of minimal observable length. Contrary to our previous
work on a non-commutative setup [14], we show that in this case,
correlations between the different modes of radiation evolve, which
reflect the fact that information emerges continuously during the
evaporation process at the quantum gravity level. This feature has
the potential to answer some questions regarding the black hole
information loss paradox [15,16] and provides a more realistic
background for treating the black hole evaporation process in its
final stages.

\section{ GUP and Parikh-Wilczek Quantum Tunneling}
The first step to discuss the quantum tunneling through the black
hole horizon is to find a proper coordinate system for the black
hole metric where the constant time slices are flat, and the
tunneling path is free of singularities. Painlev\'{e} coordinates
[17] are suitable choices in this respect. In these coordinates, the
Schwarzschild metric is given by
\begin{equation}
ds^2 = -\bigg(1 - \frac{2M}{r} \bigg) dt^2 + 2\sqrt{\frac{2M}{r}}
dtdr + dr^2 + r^2 (d\theta^2+sin^2\theta d\phi^2),
\end{equation}
which is stationary, non-static, and non-singular at the horizon.
The radial null geodesics obey
\begin{equation}
\dot{r}\equiv\frac{dr}{dt}=\pm 1-\sqrt{\frac{2M}{r}},
\end{equation}
where the plus (minus) sign corresponds to outgoing (ingoing)
geodesics. Now we incorporate the minimal length scale from quantum
gravity via the generalized uncertainty principle which motivates
modification of the standard dispersion relation [2,10].
Amelino-Camelia {\it et al} [2] (see also [18]) studied the black
hole evaporation process after an analysis of the GUP-induced
modification of the black body radiation spectrum. If GUP is
fundamental to quantum gravity, it should appear in de Broglie
relation as follows
\begin{equation}
\lambda\simeq\frac{1}{p}\left(1+\alpha L_p^2p^2\right),
\end{equation}
or
\begin{equation}
{\cal{E}}\simeq E(1+\alpha L_p^2E^2).
\end{equation}
There are other compelling reasons from noncommutative geometry and
loop quantum gravity that support relation (4) (see for instance
[2,10] and references therein). With these preliminaries, we
consider a massless particle i.e. a shell and take into
consideration the response of the background geometry to a radiated
quantum of energy $E$ with GUP correction i.e. ${\cal{E}}$. The
particle moves on the geodesics of a spacetime with $M-{\cal{E}}$
substituted for $M$. The description of the motion of particles in
the $s$-wave as spherical massless shells in a dynamical geometry
and the analysis of self-gravitating shells in Hamiltonian gravity
have been reported in Refs. [19,20](see also [21]). If one assume
that \,$t$\, increases in the direction of the future, then the
metric should be modified due to back-reaction effects. We hold the
total ADM mass, $M$, of the spacetime fixed but allow the hole mass
to fluctuate and replace $M$ by $M-{\cal{E}}$ both in the
Painlev\'{e} metric and the geodesic equation. Since the
characteristic wavelength of the radiation is always arbitrarily
small near the horizon due to the infinite blue-shift, the
wave-number approaches infinity. Therefore the WKB approximation is
valid near the horizon. In the WKB approximation, the tunneling
probability for the classically inhibited area as function of the
imaginary part of the particle action at stationary phase takes the
form
\begin{equation}
\Gamma\sim\exp(-2\textmd{Im}\, I)\approx\exp(-\beta E).
\end{equation}
As we will see later, to first order in $E$ the r.h.s of this
expression substitutes the Boltzmann factor in the canonical
ensemble characterized by the inverse temperature $\beta$. In the
$s$-wave picture, particles as spherical massless shells travel on
radial null geodesics and transfer across the horizon as outgoing
positive energy particles from $r_{in}$ to $r_{out}$. The imaginary
part of the action is thus given by
\begin{equation}
\textmd{Im}\,
I=\textmd{Im}\int_{r_{in}}^{r_{out}}p_rdr=\textmd{Im}\int_{r_{in}}^{r_{out}}\int_0^{p_r}dp'_rdr.
\end{equation}
As it is clear from the GUP expression, the commutation relation
between the radial coordinate and momentum should be modified as
follows [22]
\begin{equation}
[r,p_r]=i(1+\alpha L_p^2 p^2).
\end{equation}
In the classical limit it is replaced by the following Poisson
bracket
\begin{equation}
\{r,p_r\}=1+\alpha L_p^2 p^2.
\end{equation}
Now, using the deformed Hamilton's equation of motion,
\begin{equation}
\dot{r}=\{r,H\}=\{r,p_r\}\frac{dH}{dp_r}|_r=\left(1+\alpha L_p^2
p^2\right)\frac{dH}{dp_r}|_r
\end{equation}
where the Hamiltonian is $H=M-{\cal{E'}}$, we assume $p^2\simeq
{\cal{E'}}^2$ and eliminate the momentum in favor of energy in the
integral (6)
\begin{equation}\textmd{Im}\,
I=\textmd{Im}\int_{M}^{M-{\cal{E}}}\int_{r_{in}}^{r_{out}}\frac{(1+\alpha
L_p^2 {\cal{E'}}^2)}{\dot{r}}dr~dH=
\textmd{Im}\int_{0}^{{\cal{E}}}\int_{r_{in}}^{r_{out}}\frac{(1+\alpha
L_p^2 {\cal{E'}}^2)}
{1-\sqrt{\frac{2(M-{\cal{E'}})}{r}}}dr(-d{\cal{E'}}).
\end{equation}
The $r$ integral can be performed by deforming the contour around
the pole at the horizon, where it lies along the line of integration
and gives ($-\pi i$) times the residue
\begin{equation}
\textmd{Im}\,I=\textmd{Im}\int_{0}^{{\cal{E}}}4(-\pi
i)\left(1+\alpha L_p^2
{\cal{E'}}^2\right)(M-{\cal{E'}})(-d{\cal{E'}}).
\end{equation}
This allows us to consider the leading order correction to be just
proportional to the square of $(\sqrt{\alpha}~L_p)${\footnote{This
is the minimal length which is governed differently by the parameter
$\alpha$ in different alternatives of quantum gravity proposal. For
example, in string theory $\alpha$ takes the value which is given by
$\alpha\approx \frac{L_{p}^{2}}{\ell_{s}^{2}}$ where $\ell_{s}$ is
string length.}} for simplicity and without loss of
generality{\footnote{It can be shown easily that forthcoming results
of this paper remain credible even without this approximation.}}.
Now the imaginary part of the action takes the form
\begin{equation}
\textmd{Im}\,I=4\pi ME-2\pi E^2+\pi\alpha
L_p^2E^3\left(\frac{16}{3}M-5E\right)+O(\alpha^2L_p^4).
\end{equation}
The tunneling rate is therefore
\begin{equation}
\Gamma\sim\exp\left(-8\pi ME+4\pi E^2-2\pi\alpha
L_p^2E^3\left(\frac{16}{3}M-5E\right)+O(\alpha^2L_p^4)\right)=\exp(\Delta
S),
\end{equation}
where $\Delta S$ is the difference in black hole entropies before
and after emission [13,23]. In string theory it is anticipated that
the tunneling rates from excited D-branes in the microcanonical
ensemble depends on the final and initial number of microstates
available to the system. In a more precise expression, it was shown
that the emission rates on the high-energy scale correspond to
differences between counting of states in the microcanonical and
canonical ensembles [23]. The first and second expressions in the
exponential exhibit a similar kind of non-thermal aberration that
was discovered in Ref. [13]. So, the emission spectrum cannot be
strictly thermal. In our situation, there is an additional term
depending on the GUP parameter in first order that cannot be
neglected once the black hole mass becomes comparable to the Planck mass.\\
We now illustrate whether or not, the emission rates for the
different modes of radiation during the evaporation are mutually
related from a statistical viewpoint. Utilizing (13), the emission
rate for a first quantum with energy $E_1$, gives
\begin{equation}
\ln\Gamma_{E_1}=-8\pi ME_1+4\pi E_1^2-2\pi\alpha
L_p^2E_1^3\left(\frac{16}{3}M-5E_1\right).
\end{equation}
Similarly, the emission rate for a second quantum $E_2$, takes the
form
\begin{equation}
\ln\Gamma_{E_2}=-8\pi (M-E_1)E_2+4\pi E_2^2-2\pi\alpha
L_p^2E_2^3\left(\frac{16}{3}(M-E_1)-5E_2\right).
\end{equation}
Correspondingly, the emission rate for a single quantum with the
same total energy, $E=E_1+E_2$, yields
\begin{equation}
\ln\Gamma_{(E_1+E_2)}=-8\pi M(E_1+E_2)+4\pi (E_1+E_2)^2-2\pi\alpha
L_p^2(E_1+E_2)^3\left(\frac{16}{3}M-5(E_1+E_2)\right).
\end{equation}
It can be approved that these probabilities are correlated. On the
other hand, the non-zero statistical correlation function is
\begin{equation}
\chi(E_1+E_2;E_1,E_2)=4\pi\alpha L_p^2
E_1E_2\left(-8M(E_1+E_2)+10E_1^2+15E_1E_2+\frac{22}{3}E_2^2\right).
\end{equation}
This means that not only the probability of tunneling of two
particles of energy $E_1$ and $E_2$ is not similar to the
probability of tunneling of one particle with their compound
energies, $E_1+E_2$, but there are also correlations between them.
In fact, whenever one quantum of emission is radiated from the
surface of the black hole horizon, the aberrations are created on
the Planck scale that influence the second quantum of emission, and
these aberrations cannot be neglected, particularly once the black
hole mass becomes comparable with the Planck mass. Therefore, as
expected in Ref. [24], in this way the form of the amendments as
back-reaction effects with incorporation of GUP influences are
sufficient by themselves to recover information. Information leaks
out from the black hole as the non-thermal GUP correlations within
the Hawking radiation.

Currently there are four main proposals about what happens to the
information that falls into a black hole. First proposal is that the
black hole can evaporate completely as uncorrelated thermal
radiation in each mode, and all the information including the
original quantum state that formed the black hole, excluding its
mass, charge and angular momentum, would disappear from our
universe. This means that, in particular, this proposal causes
allowing pure states to evolve into mixed states, incompatibility
with the basic principles of quantum mechanics. Second proposal is
that the black hole can completely disappear, but the information
appears in the final burst of radiation when the black hole shrinks
to the Planck size. A third possibility is that the black hole never
disappears completely, and the information is not lost, but would be
stored in a Planck size stable remnant. And fourth idea is that the
information comes out in non-thermal correlations within the Hawking
radiation, the process being portrayed by a unitary $S$-matrix. In
other words, there are non-thermal correlations between different
modes of radiation throughout the evaporation process that
information emerges ceaselessly encoded through them. In this paper,
we have studied the credibility of the fourth conjecture within a
perturbative quantum gravity approach. We have shown that equation
(17) represents the recovery of the information as real correlations
at the Planck scale, where quantum gravity corrections become
important.

In summary, the inclusion of the effects of quantum gravity as
modification of the de Broglie relation and corresponding
commutation relations in the quantum tunneling framework of the
black hole evaporation, leads to correlation between emitted modes
of evaporation. In this setup, information leaks out of the black
hole in the form of non-thermal GUP correlations, which might solve
the black hole information loss
paradox.\\

{\bf Acknowledgment}\\
We would like to thank Elias Vagenas for useful comments. We would
also like to thanks Professor Rudolf Treumann and three anonymous
referees for important contribution in this work.\\
This work has been supported partially by Research Institute for
Astronomy and Astrophysics of Maragha, Iran.

\end{document}